\documentclass[aps,prd,preprintnumbers,twocolumn,groupedaddress,nofootinbib,showpacs]{revtex4}
\usepackage{graphicx}
\usepackage{amsmath}
\usepackage{amssymb}
\usepackage{array}
\usepackage{slashed}
\usepackage{undertilde}
\usepackage{subfigure}
\usepackage{float} 
\usepackage{color}
\definecolor{lightred}{rgb}{0.7,0.1,0.1}

\def\lsim{\mathrel{\raise.3ex\hbox{$<$\kern-.75em\lower1ex\hbox{$\sim$}}}}
\def\gsim{\mathrel{\raise.3ex\hbox{$>$\kern-.75em\lower1ex\hbox{$\sim$}}}}

\newcommand{\met}{\slashed{E}_T}
\newcommand{\GeV}{\text{GeV}}
\newcommand{\Qtr}{Q_\text{tr}}

\begin{document}

\title{Mono-Higgs Detection of Dark Matter at the LHC}
\author{Asher Berlin}
\author{Tongyan Lin}
\author{Lian-Tao Wang}
\affiliation{Enrico Fermi Institute and Kavli Institute for Cosmological Physics, \\ The University of Chicago, Chicago, Illinois, 60637-1433}
\date{\today}

\begin{abstract}
Motivated by the recent discovery of the Higgs boson, we investigate
the possibility that a missing energy plus Higgs final state is the
dominant signal channel for dark matter at the LHC. We consider
examples of higher-dimension operators where a Higgs and dark matter
pair are produced through an off-shell $Z$ or $\gamma$, finding
potential sensitivity at the LHC to cutoff scales of around a few
hundred GeV. We generalize this production mechanism to a simplified
model by introducing a $Z'$ as well as a second Higgs
doublet, where the pseudoscalar couples to dark matter. 
Resonant production of the $Z'$ which decays to a Higgs
plus invisible particles gives rise to a potential mono-Higgs signal.
This may be observable at the 14 TeV LHC at low $\tan \beta$ and when
the $Z'$ mass is roughly in the range 600 GeV to 1.3 TeV.
\end{abstract}

\maketitle

\section{Introduction}
Dark matter (DM) contributes a large component of the mass-energy of
the universe. The leading hypothesis is that most of the dark matter is
in the form of stable, electrically neutral, massive particles, which
interact at least gravitationally with baryons. If such a 
particle interacts non-gravitationally with standard model (SM)
particles as well, for instance via the weak force, detecting it
through high-energy collisions at particle accelerators is one of the
most promising avenues towards identifying the specific nature of its
detailed interactions. For instance, if DM production is kinematically
accessible at the Large Hadron Collider (LHC), then missing energy
signatures that deviate from SM predictions would provide compelling
evidence of new stable, electrically neutral particles, and thus
strong candidates for cosmological DM.

Various approaches to describing particle DM interactions have been
explored in order to understand possible detection signatures at the
LHC. The most detailed set of attempts include complete quantum field
theories incorporating many new particles into the SM, for example
supersymmetric dark matter \cite{Jungman:1995df}. Such top-down, or
UV-complete, theories often have large sets of a priori undetermined
additional parameters. Therefore, making confident phenomenological
predictions can become very burdensome.

On the opposite end of the spectrum, one can assume that aside from
the dark matter, any new heavy fields of the UV-complete theory can be
integrated out and that the relevant physics can be described by an
effective field theory (EFT). The effective field theorist thus
proceeds by writing down a tower of non-renormalizable contact
operators governing the DM interactions with SM particles. The
underlying UV theory determines the coefficients of these operators,
which in turn can be constrained in a model-independent way from
experimental results and also be related in a simple way to relic
density or direct detection predictions \cite{Beltran:2010ww}. In
addition to the relative simplicity of constraining individual
operators, this approach has a particular appeal at a time when no
other signs of new physics have yet been discovered at the LHC.

Recent studies taking advantage of the EFT technique have considered
collider signals such as monophoton \cite{Fox:2011fx}, monojet
\cite{Fox:2011pm,Rajaraman:2011wf,ATLAS:2012ky,Chatrchyan:2012me},
mono-$Z/W$
\cite{Carpenter:2012rg,Aad:2013oja} and
mono-$b$ events \cite{Lin:2013sca}, during which one (or more)
particle of the SM is produced and detected in the collider, recoiling
against some missing transverse energy (MET or $\slashed{p}_T$)
associated with the DM. This work has been generalized to a set of
so-called ``simplified models'' where the DM couples to the SM through
renormalizable interactions, for example through a new mediator that
can be produced on-shell
\cite{An:2012va,Bai:2013iqa,Chang:2013oia,DiFranzo:2013vra,Papucci:2014iwa}.

\begin{figure*}[tbh]
\mbox{\includegraphics[width=0.47\textwidth,clip]{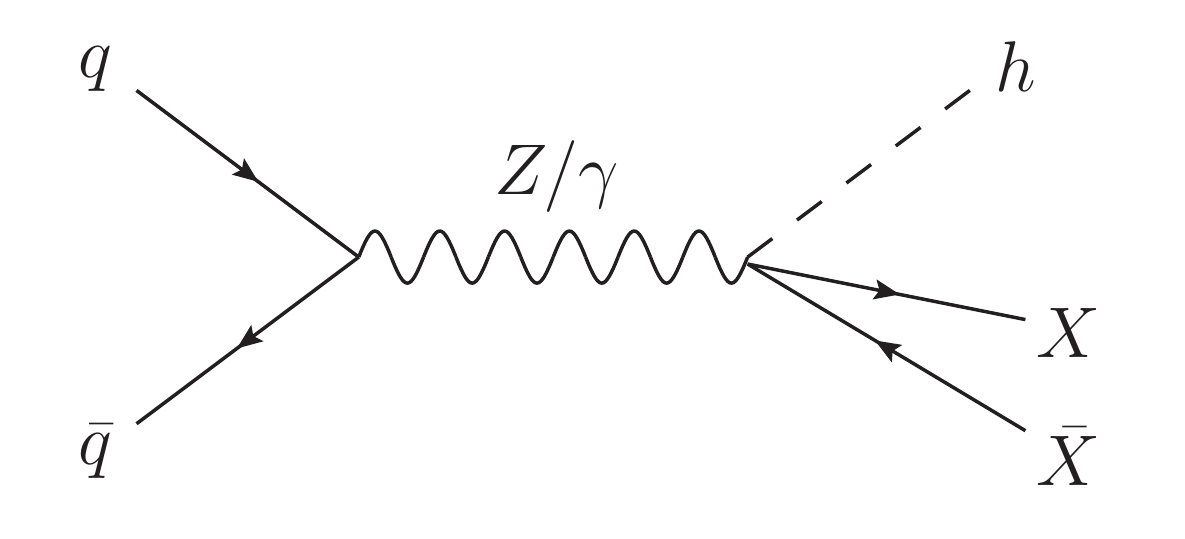}} 
\mbox{\includegraphics[width=0.47\textwidth,clip]{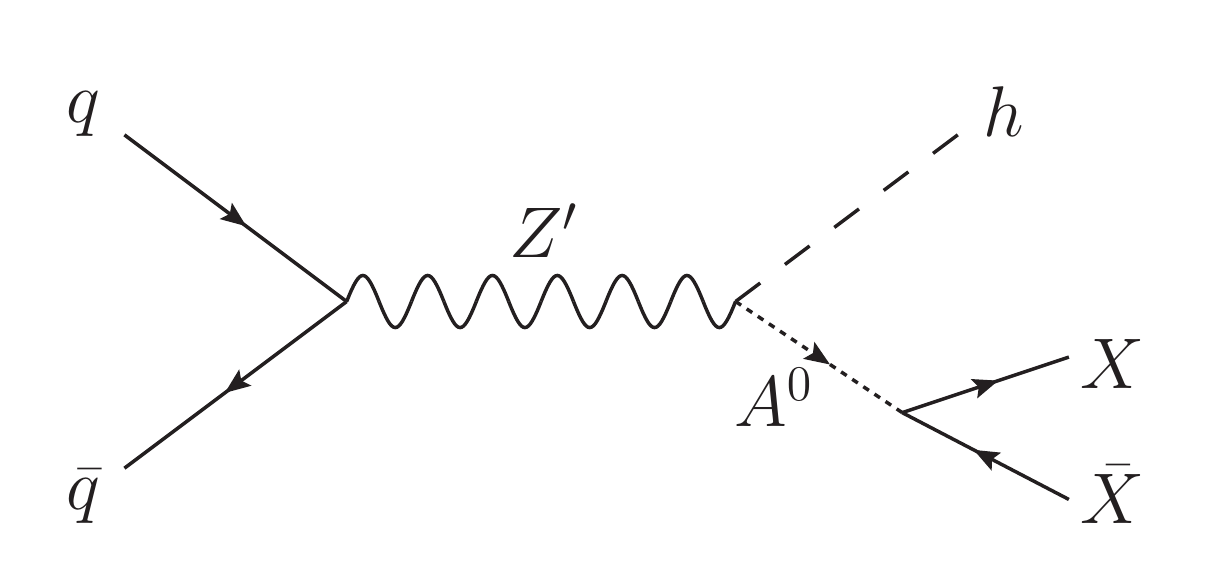}} 
\caption{Production mechanisms for dark matter plus Higgs through
  ({\it left}) a contact operator coupling dark matter to $Zh$ or
  $\gamma h$, or ({\it right}) a new $Z'$ coupled to a two Higgs
  doublet model, where the new pseudoscalar $A^0$ decays primarily to
  the dark matter.}
\label{fig:diagrams}
\end{figure*}

In light of the recent Higgs discovery at the LHC
\cite{Aad:2012tfa,Chatrchyan:2012ufa}, we can expand our search in yet
another avenue. In this paper we investigate the possible production
of a Higgs along with DM, which is accordingly dubbed a mono-Higgs
process. The observed final states are MET plus the Higgs decay
products, with an invariant mass constrained to be relatively close to
the true mass $m_h \approx 125$ GeV.  

The focus of our article is to explore those possibilities where
mono-Higgs could be the primary production mechanism for DM at the
LHC. We will consider examples of both contact operators and
simplified models. We begin in Section~\ref{sec:LHC} with a discussion
of LHC searches for Higgs plus MET final states, concentrating here on
the $b\bar b$ and diphoton decay channels for the Higgs. In
Section~\ref{sec:EFT}, we present examples of higher-dimension
operators coupling DM to Higgs doublets and electroweak gauge
bosons. We derive constraints on the coefficients of these operators
both with and without implementing a unitarity condition on the
potential signal events. Motivated by the processes in the EFT
description, in Section~\ref{sec:ZPrime} we introduce a simplified
model with a $Z'$ gauge boson and two Higgs doublets, where the dark
matter is coupled to the heavy pseudoscalar Higgs. We demonstrate that
the 14 TeV LHC can probe the parameter space of this model at low
$\tan \beta$. We conclude in Section~\ref{conc}.

We also note that the mono-Higgs signal has recently been discussed in
Refs.~\cite{Petrov:2013nia,Carpenter:2013xra}. Ref.~\cite{Petrov:2013nia}
considered contact operators coupling dark matter to SM Higgs doublets
and possibly other SM states (the operators are different from the
ones in this paper); however they found that for most of the operators
the bounds on the cutoff scale are quite low, less than 50 GeV, which
is well beyond the regime of validity for assuming a contact operator.

Ref.~\cite{Carpenter:2013xra} considered a somewhat different set of
operators as well as simplified models. For the ``Higgs-portal''-type
operators ({\it e.g.},
\cite{Burgess:2000yq,Djouadi:2011aa,Greljo:2013wja}), they find LHC
limits to be much weaker than exclusion limits on Higgs invisible
decay for DM masses below $m_h/2$, while direct detection is very
constraining at higher masses.
Ref.~\cite{Carpenter:2013xra} also considered simplified
models with an additional $Z'$, where the Higgs is produced through
Higgs-strahlung of the $Z'$. For the case of $Z-Z'$ mass mixing, they
found mono-Higgs is only able to probe large mixing angles ($\sin
\theta > 0.1$), in apparent conflict with precision electroweak data.
In contrast, for our scenario the $Z'$ is produced resonantly and
decays, and we have imposed the precision electroweak constraint from
fits of the $\rho_0$ parameter.

\section{$\text{Higgs} + \text{MET}$ at the LHC}
\label{sec:LHC}

We consider two possible Higgs decay channels, $b \bar b$ and $\gamma
\gamma$, as promising for observing Higgs plus MET.  The $b\bar b$
channel has the largest branching ratio for a Higgs of mass $m_h =
125$ GeV, $\text{Br}(h \to b \bar b) \approx 0.577$
\cite{Heinemeyer:2013tqa}, and gives the best statistics for the
signal, while the diphoton branching ratio is only $\text{Br}(h \to
\gamma \gamma) \approx 2.28~\times~10^{-3}$, but is potentially a very
clean channel. These channels as well as multi-lepton final states
from $h \to Z Z^*$ were also studied in \cite{Carpenter:2013xra}.

The dominant irreducible SM background for Higgs plus MET is $Zh$
production with $Z$ decaying to neutrinos. Depending on the decay
channel, other SM backgrounds can also be comparable or larger. Here
we rely on the ATLAS report \cite{ATLAS-CONF-2013-079} to derive
bounds from LHC Run 1. For 14 TeV projections, we estimate backgrounds
rates from our own Monte Carlo event simulations and also use some
results from \cite{Carpenter:2013xra}.

Our dark matter models have been implemented with FeynRules 2.0
\cite{Alloul:2013bka}, and our event generation makes use of the
MadGraph \cite{Alwall:2011uj}, PYTHIA \cite{Sjostrand:2006za}, and Delphes
\cite{deFavereau:2013fsa} pipeline from parton-level to detector-level
simulation.

\renewcommand\arraystretch{1.4}
\begin{table}[tb]
\begin{center}
\begin{tabular}{|c|c|c|c|c|}
\hline
  & LHC Run 1 &  14 TeV \\
\hline
$t \bar t$  &  200  &  $1006 \pm 335$ \\
$Z b \bar b$      &  336  &  $682 \pm 26$ \\
$Vh$	&  23   &   $142 \pm 5$   \\
SM total    &  $727 \pm 11$  &  $1830 \pm 336$ \\
\hline
{\small Dim-8, fermion DM} &   $329 \pm 10$ &  $23150 \pm 880$ \\
$M_{Z'}=1$ TeV, $\tan \beta = 1$ &  $43 \pm 1$ &  $1836 \pm 36$ \\
\hline
\end{tabular}
\end{center} 
\caption{Background and signal events for $h \to b \bar b$ decay, for
  the cuts described in the text. The background numbers for LHC Run 1
  are taken from Ref.~\cite{ATLAS-CONF-2013-079} for MET $>$ 120 GeV.
  For our background estimate at a 14 TeV LHC, we include only the
  processes listed here; uncertainties from MC statistics are shown
  and we include an additional 25$\%$ systematic uncertainty in
  deriving constraints.  For the signal from a dimension-8 operator
  with fermion DM, Eq.~(\ref{eq:dim8fermion}), we take fiducial values
  of $\Lambda = 200$ GeV and $m_X = 1$ GeV. For the $Z'$ case, the
  coupling is the upper limit allowed by the $\rho_0$ constraint,
  shown in Fig.~\ref{fig:gzlimits}.
\label{tab:14TeV}}
\end{table}
\renewcommand\arraystretch{1.0}

\subsection{Two b-jet channel \label{sec:bbchannel}}

A search for $h \rightarrow b \bar{b}$ decay in association with a
$Z/W$ boson has been performed using the data of Run 1 of the LHC; the
observed signal strength is compatible with that of the SM Higgs boson
\cite{ATLAS-CONF-2013-079,Chatrchyan:2013zna}. In particular, the
ATLAS collaboration presents an analysis for the $Z(\nu \bar \nu)h$
channel in several MET bins, with the full integrated luminosity of
$4.7$/fb at 7 TeV and 20.3/fb at $\sqrt{s} = 8$ TeV
\cite{ATLAS-CONF-2013-079}. We use these results to derive constraints
on mono-Higgs for the models in this paper.

Event selection is governed by demanding two leading $b$-tagged jets,
with $p_T > 20$ GeV and $|\eta| < 2.5$, with the highest $p_T$
$b$-tagged jet having $p_T > 45$ GeV. Multijet backgrounds are reduced
by requiring $\met > 120$ GeV and constraints on the azimuthal angle
between the missing transverse momentum and jets: $\Delta \phi(\met, b
\bar b) > 2.8, {\rm min}[\Delta \phi(\met, j)] > 1.5$. A lepton veto
is imposed, and the $b \bar b$ system invariant mass must reconstruct
to near the Higgs mass, 90 GeV $< m_{bb} < 150$ GeV. Finally, $t \bar
t$ is suppressed by vetoing events that have any additional jets with
$p_T > 30$ GeV.

Estimates of SM processes, including $Zh$, are compared to observed
data events in three MET bins. The most important backgrounds are
$Z+b \bar b$ and $t\bar t$. Making use of these published SM process
estimates, we compare our signal to the data with cuts of $\met >
120\ \GeV$, $\met > 160\ \GeV$, and $\met > 200\ \GeV$, and derive
95$\%$ CL upper limits on the number of possible mono-Higgs signal
events.

For 14 TeV projections, we modify the 8 TeV ATLAS cuts slightly,
loosing the jet veto such that up to one additional jet with $p_T >
30$ GeV is allowed, and take a cut of $\met > 250$ GeV. The total
integrated luminosity is 300/fb. Our estimates for background rates
are shown in Table~\ref{tab:14TeV}. We find the $b \bar b$ channel
performs better compared to the results in
Ref.~\cite{Carpenter:2013xra}; this appears to be due primarily to our
choice of $R=0.4$ jet clustering radius instead of $R=0.7$, since with
a larger radius the two $b$-jets from the Higgs decay are more often
clustered together in the boosted Higgs regime.


\begin{figure*}[thb]
\mbox{\includegraphics[width=0.48\textwidth,clip]{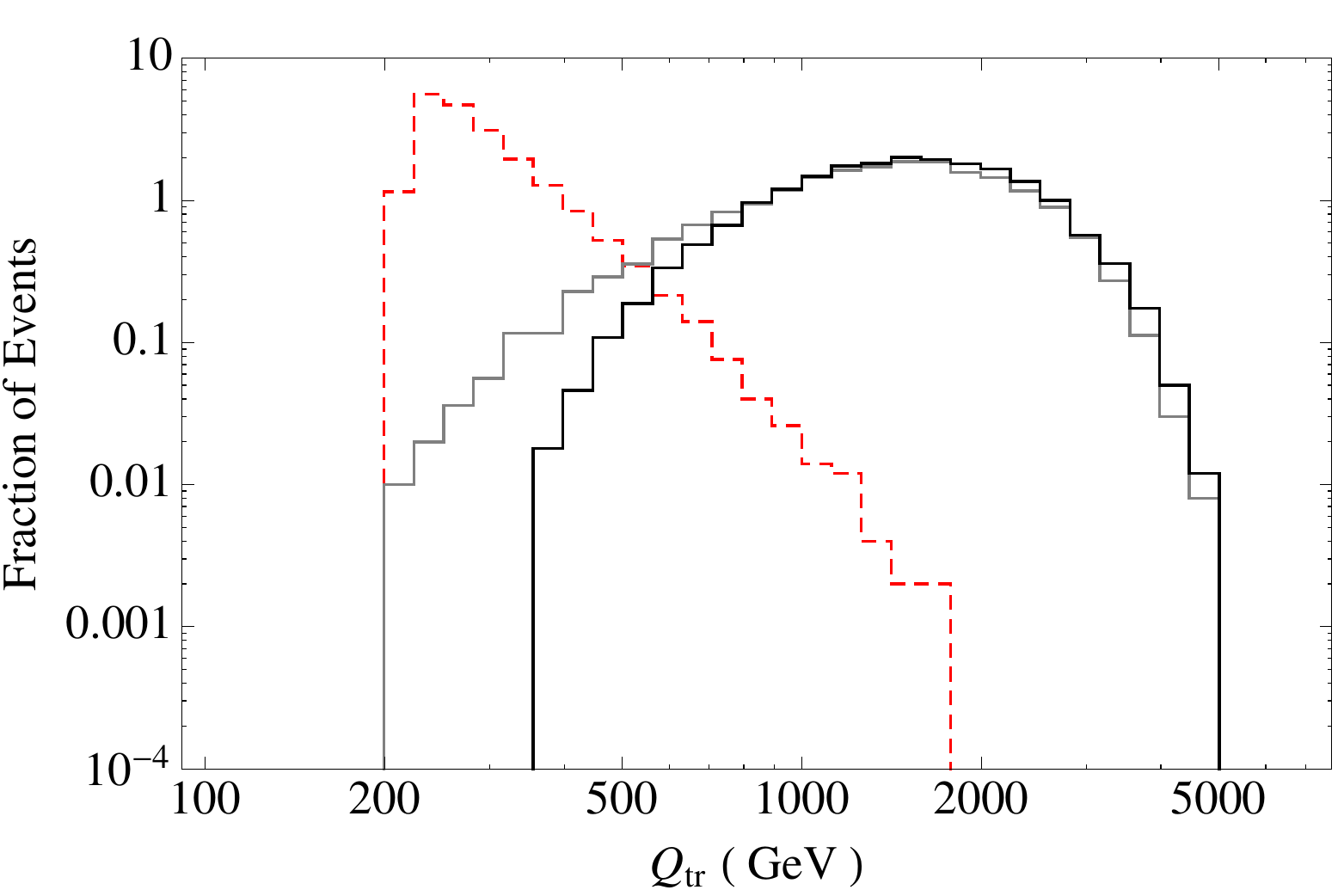}} 
\mbox{\includegraphics[width=0.48\textwidth,clip]{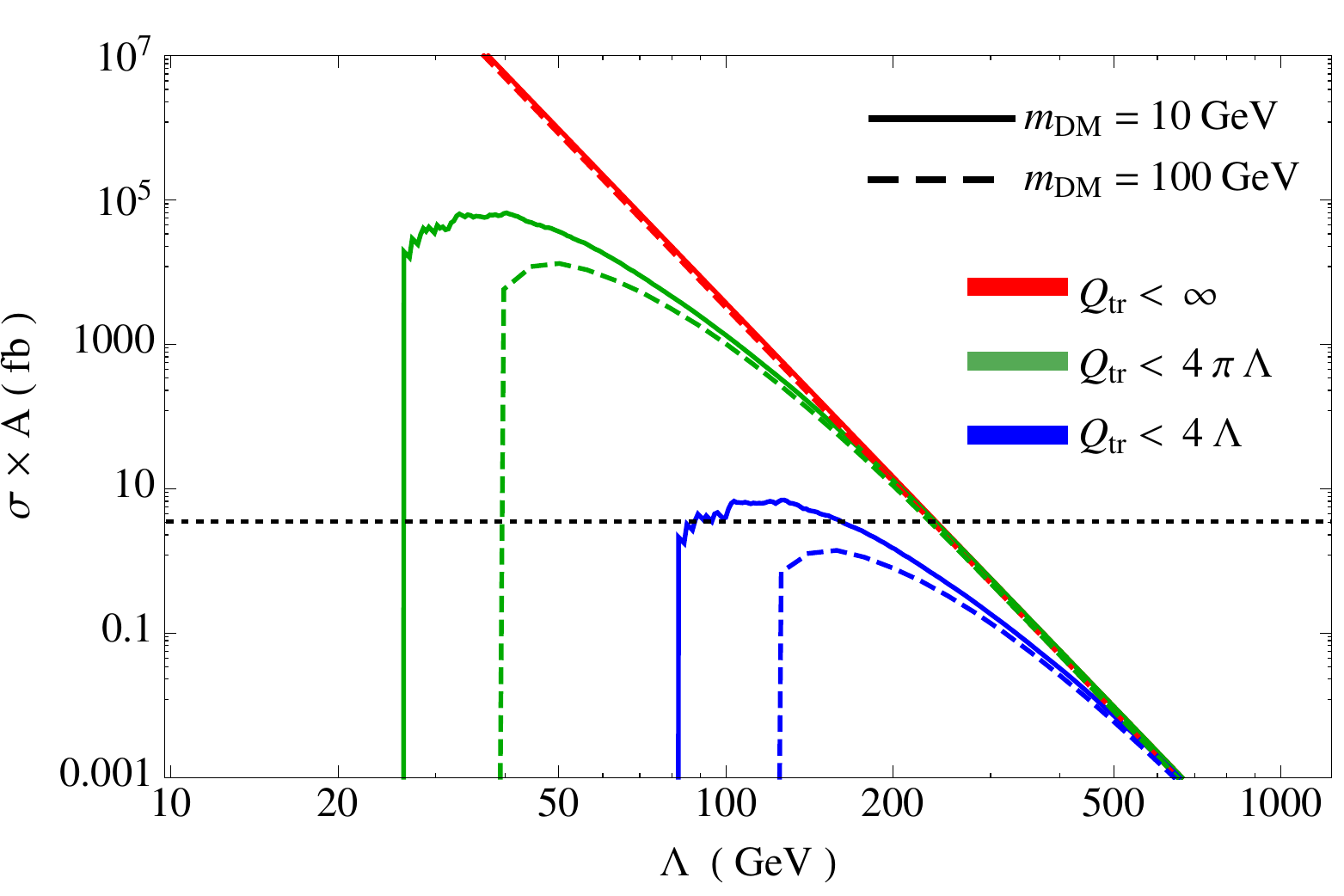}} 
\caption{ \label{fig:eft} ({\it Left}) Distributions (at $\sqrt{s}= 8$
  TeV) for the momentum transfer $Q_{\text{tr}}$ for 10 GeV (solid
  gray line) and 100 GeV (solid black line) DM, from a mono-Higgs DM
  signal corresponding to the operator of Eq.~(\ref{eq:dim8fermion})
  with fermion DM. The irreducible SM background from $Zh$ production
  (red dashed line) is also shown. ({\it Right}) For the same
  operator, the rate for mono-Higgs at 8 TeV with a cut of 120 GeV
  missing transverse energy.  The total cross section is scaled by the
  fraction of events satisfying various ``unitarity'' conditions on
  $\Qtr$. The horizontal line indicates the approximate cross section
  that would be ruled out at 95$\%$ CL using data from Run 1 of the
  LHC; regions of $\Lambda$ with cross sections above this line are
  excluded and correspond to the shaded regions in
  Fig.~\ref{fig:eftbounds}.}
\end{figure*}

\subsection{Diphoton channel \label{sec:aachannel}}

The diphoton channel requires two hard photons reconstructing to the
Higgs mass, large missing energy, and a veto on leptons. The dominant
SM backgrounds are $Z\gamma\gamma$ and $hZ/hW$.  Because the Higgs
branching ratio to photons is so small, we find that this channel is
not constraining if the 8 TeV run is considered, since there are
simply not enough signal events. However, the statistics are far
improved at 14 TeV.  We use results for background estimates from
\cite{Carpenter:2013xra}, where they found that this channel can
demonstrate improved sensitivity over $b \bar b$ (which suffers from a
larger $t \bar t$ background). The cuts applied require
$m_{\gamma\gamma} \in [110,130]$ GeV and $\met > 100,250$ GeV at 8,14
TeV respectively.

\section{Effective Field Theory \label{sec:EFT}}

\begin{figure*}[tbh]
\mbox{\includegraphics[width=0.48\textwidth,clip]{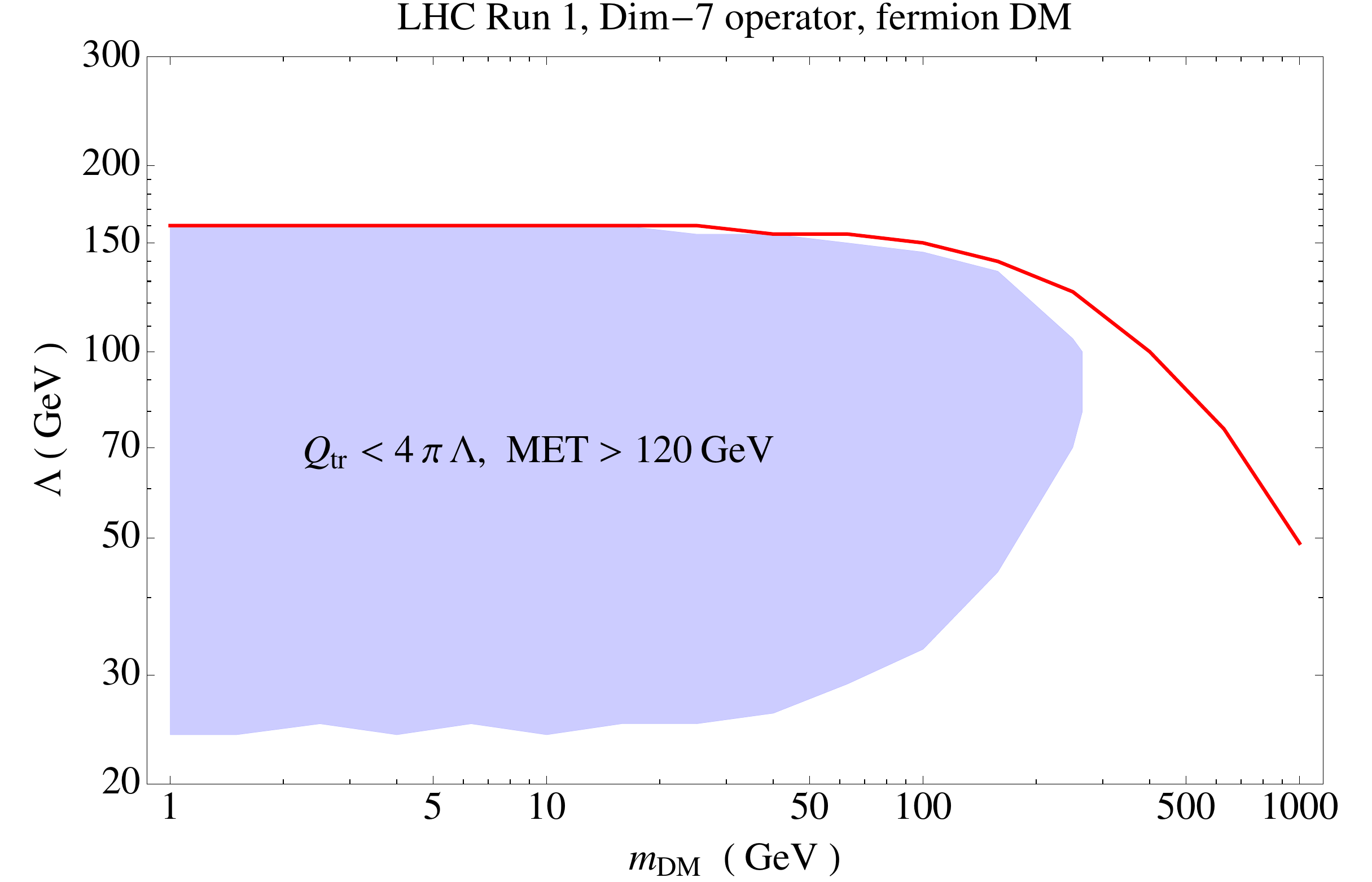}} 
\mbox{\includegraphics[width=0.48\textwidth,clip]{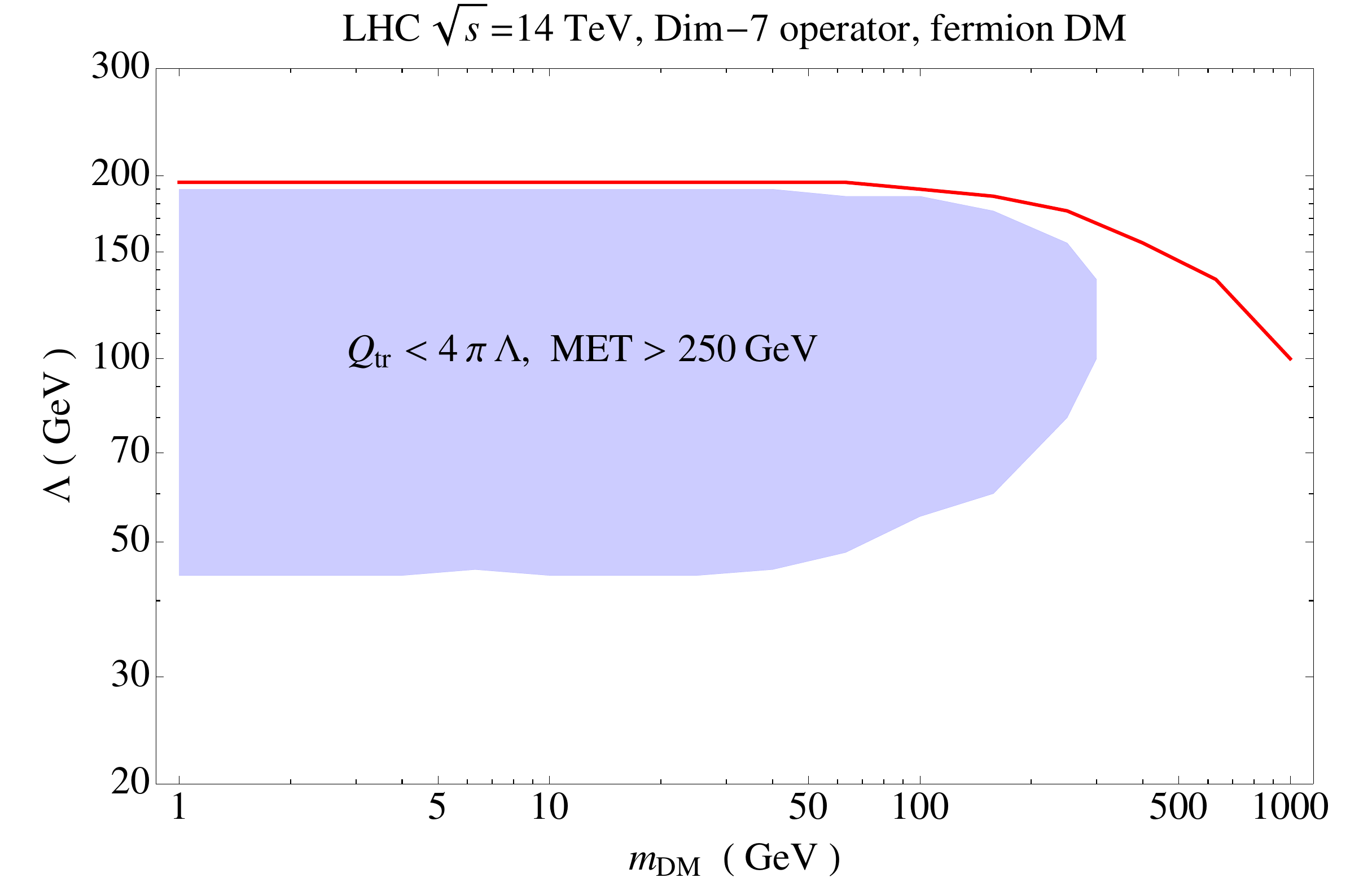}}
\mbox{\includegraphics[width=0.48\textwidth,clip]{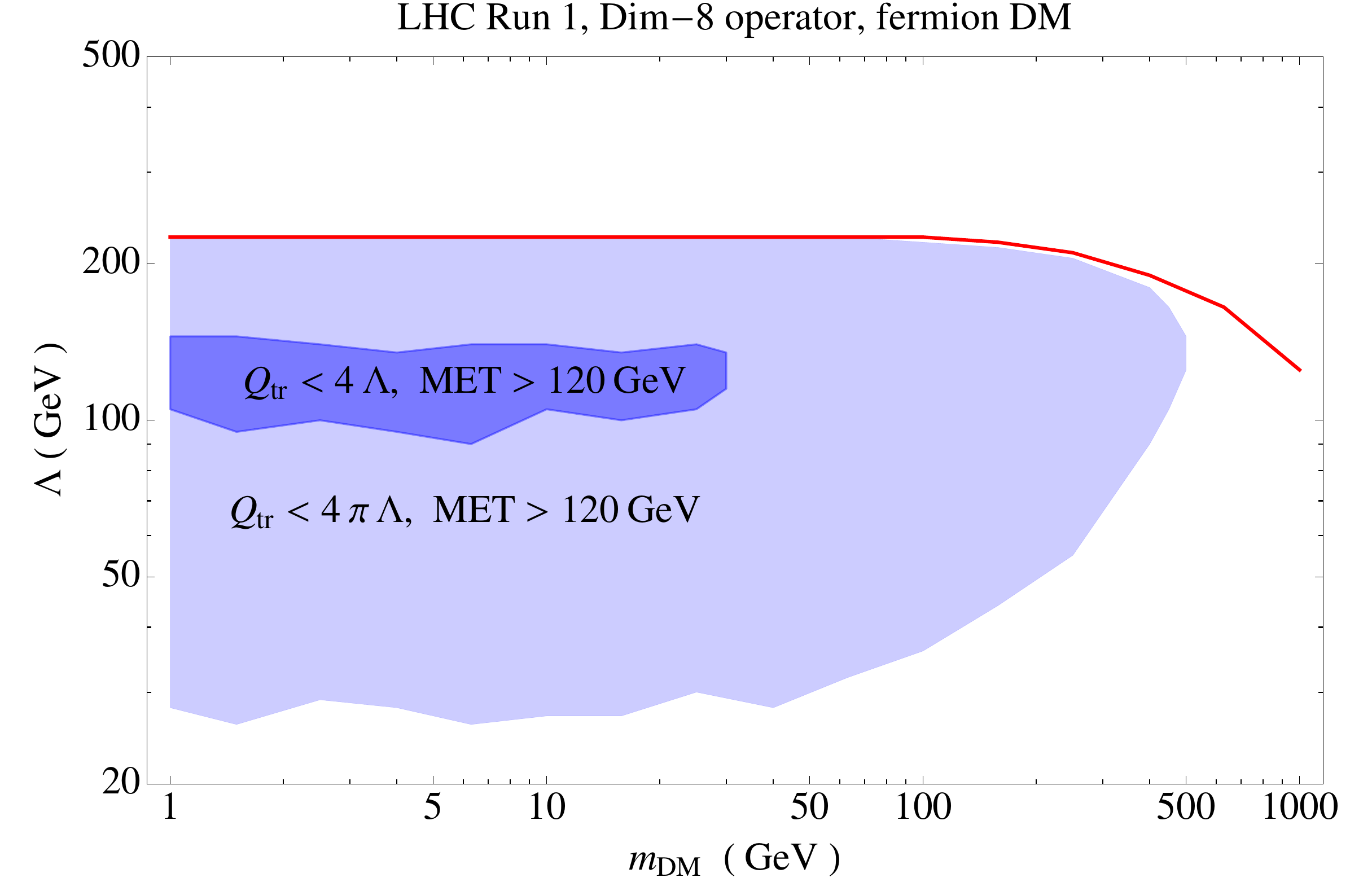}} 
\mbox{\includegraphics[width=0.48\textwidth,clip]{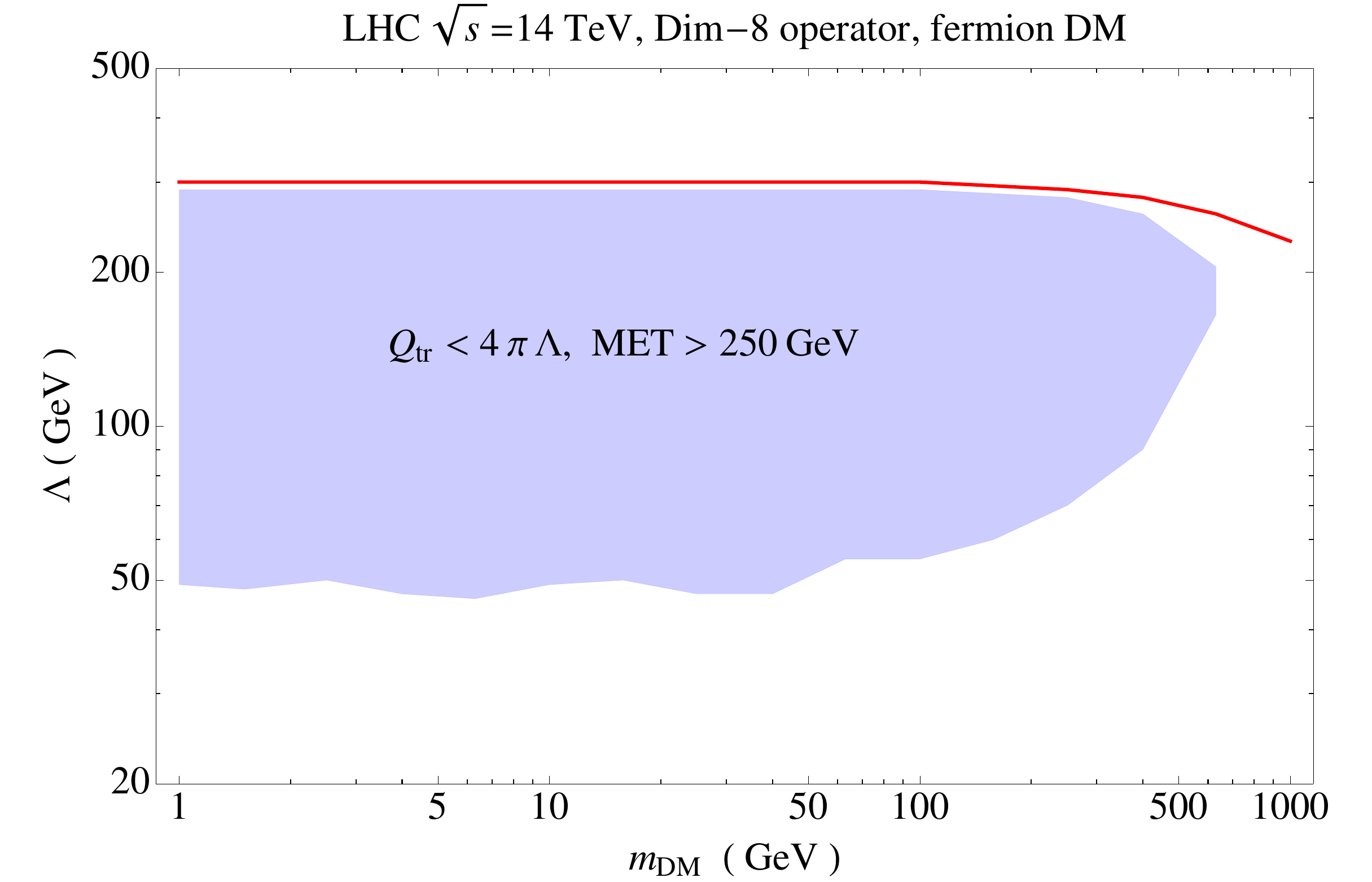}} 
\mbox{\includegraphics[width=0.48\textwidth,clip]{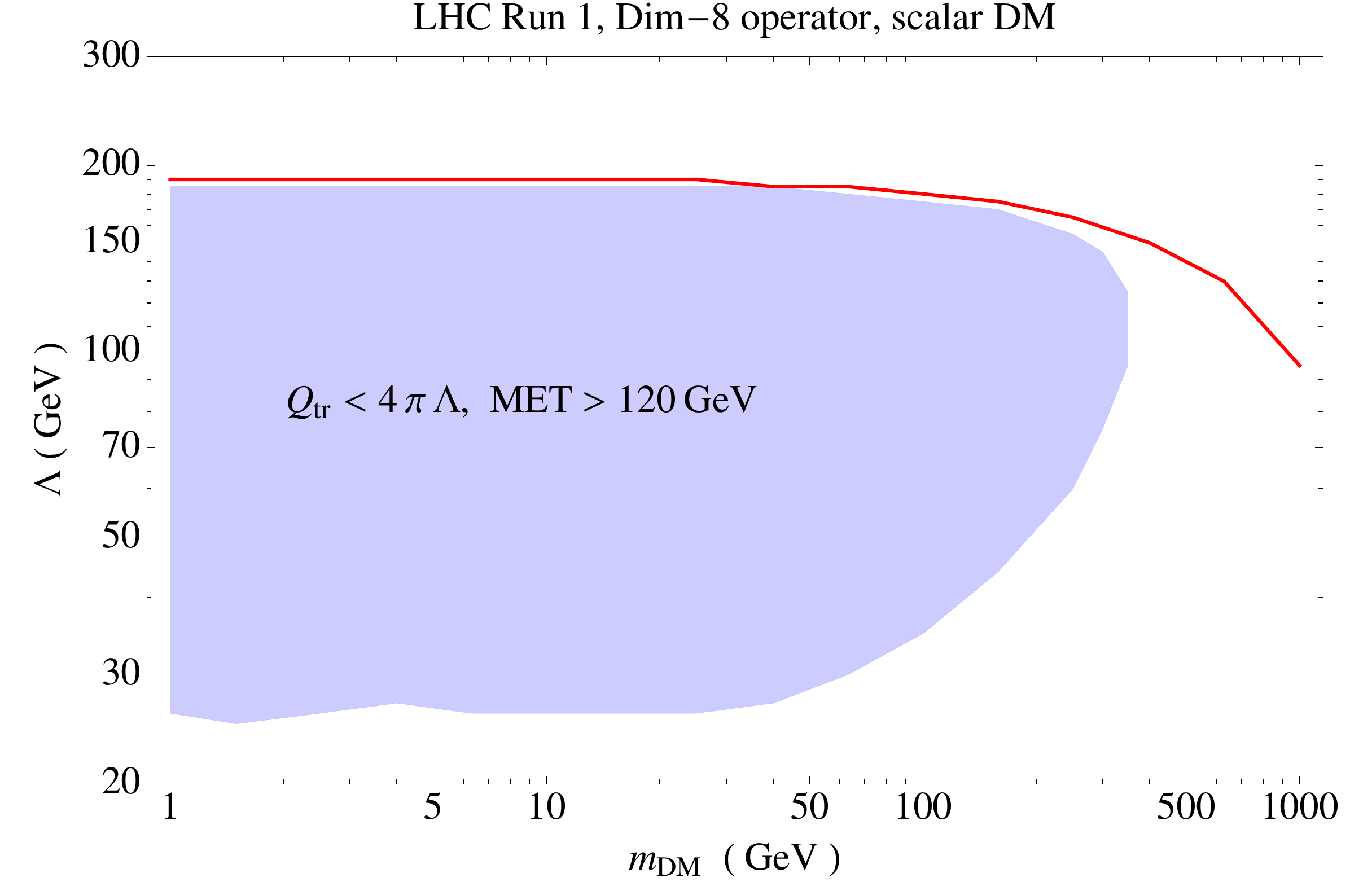}}  
\mbox{\includegraphics[width=0.48\textwidth,clip]{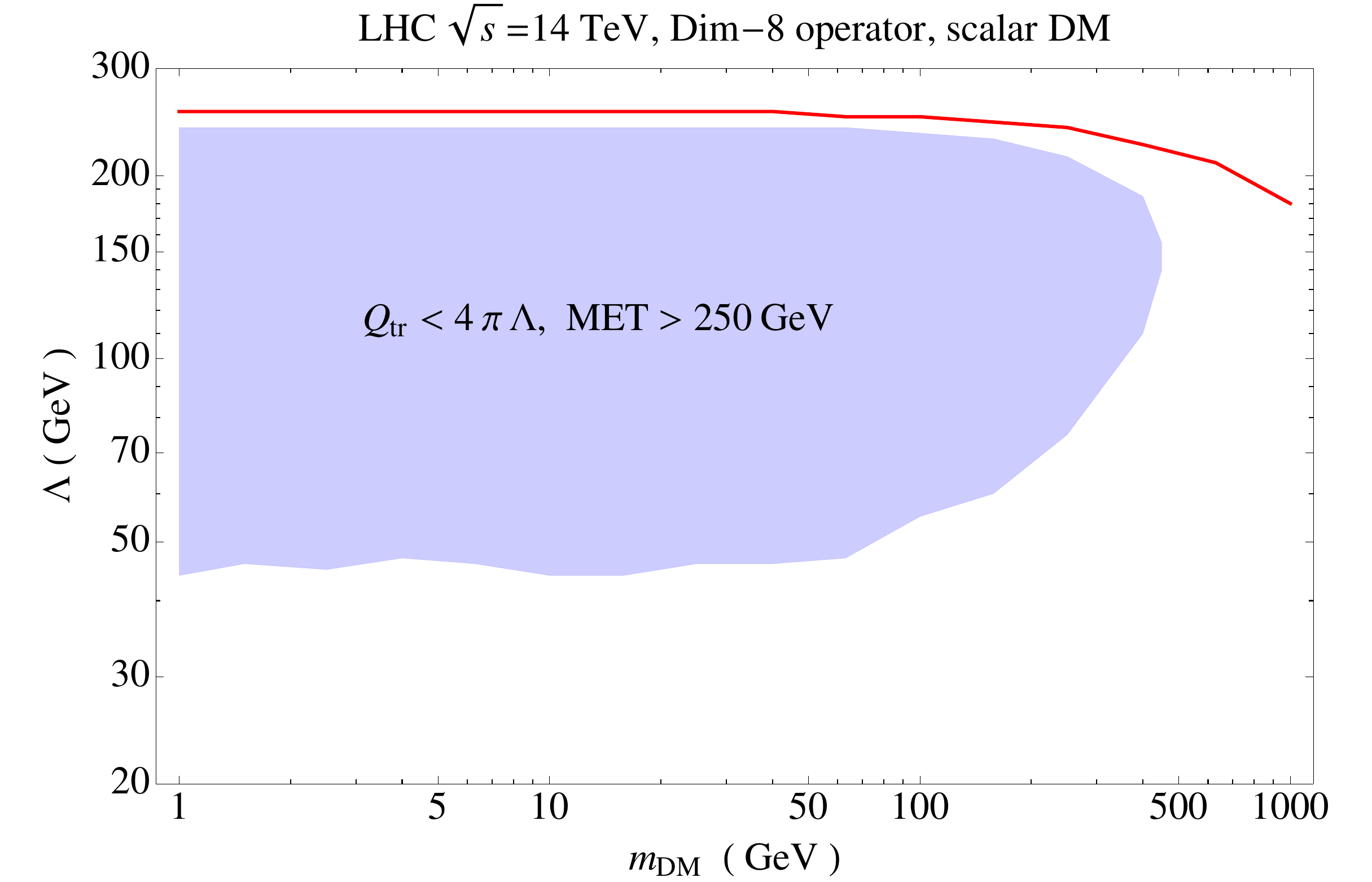}} 
\caption{95$\%$ CL constraints from mono-Higgs on the suppression
  scale $\Lambda$ as a function of DM mass, for operators discussed in
  Section~\ref{sec:EFT}. The dimension-7 operator is
  $\frac{i}{\Lambda^3}\bar X \gamma^{\mu\nu} X \left[ (D_\mu
    H)^\dagger D_\nu H - \text{h.c.}  \right]$, the dimension-8
  operator coupling to fermion DM is $\frac{1}{\Lambda^4} \bar{X}
  \gamma^\mu X \ ( W^a_{\nu \mu} H^\dagger t^a D^\nu H + \text{h.c.}
  )$, and the dimension-8 operator coupling to scalar DM is
  $\frac{1}{\Lambda^4} \frac{1}{2} ( \phi^\dagger \partial^\mu \phi +
  \text{h.c.} ) \ ( B_{\nu \mu} H^\dagger D^\nu H + \text{h.c.} )$.
  The solid lines are the lower bounds for the naive EFT result
  ($Q_{\text{tr}} < \infty$). The shaded regions are the excluded
  regions imposing the conditions on the momentum transfer
  $Q_{\text{tr}} < 4 \pi \times \Lambda$ or $Q_{\text{tr}} < 4
  \Lambda$ to address the apparent violation of unitarity. The left
  column shows LHC Run 1 limits, derived for the $b \bar b$ channel
  with a MET cut of 120 GeV, while right column shows 14 TeV limits
  assuming the diphoton channel and a MET cut of 250 GeV.
\label{fig:eftbounds}}
\end{figure*}


Contact operators coupling dark matter to a Higgs doublet can
potentially give rise to a mono-Higgs signal. If the dark matter is a
gauge singlet, then gauge invariance implies the operator must also
include other electroweak doublets. We focus on operators that give
rise to a coupling of dark matter to both $h$ and $Z/\gamma$, allowing
the production of dark matter through the process shown in
Fig.~\ref{fig:diagrams}.  If the dark matter couples to two Higgs
bosons, the production rate is correspondingly lower.

For the process above, we also note that a mono-$Z$ signal is possible
by reversing the roles of the $h$ and the $Z$; this rate is
automatically lower by several orders of magnitude since it requires
the initial production of an $s$-channel Higgs. For all the operators
considered here, the limits from mono-$Z$ are weaker
compared to mono-Higgs.

These kinds of operators have been studied in
Refs.~\cite{Chen:2013gya,Fedderke:2013pbc,Carpenter:2013xra}, as well
as mono-Higgs from Higgs-portal type operators in
\cite{Petrov:2013nia}. The lowest dimension SM operator that can give
a $Zh$ interaction with dark matter is
\begin{equation}
  i(H^\dagger D_\mu H - \text{h.c.})  \rightarrow  -2 m_Z h Z_\mu - \langle v \rangle m_Z Z_\mu,
  \label{eq:SMdim3}
\end{equation}
after electroweak symmetry breaking. This operator could be combined
with singlets formed of the dark matter: $i(\phi^\dagger \partial^\mu
\phi - \text{h.c.})$ for scalar DM, and $\bar X \gamma^\mu X$ or $\bar
X \gamma^\mu \gamma^5 X$ for fermion DM.  Because of the induced
direct $Z$ coupling to dark matter, direct detection is very
constraining for $m_{\text{DM}} > 10\ \GeV$, while the invisible $Z$
width is very constraining for $m_{\text{DM}} < m_Z/2$.  Despite this,
in the case of scalar DM Ref.~\cite{Carpenter:2013xra} found that a
mono-Higgs search at 14 TeV could be much more sensitive than the
invisible $Z$ width\footnote{The constraints on the suppression scale
  $\Lambda$ are again of order a few hundred GeV up to a TeV for the
  LHC and therefore has the same problem with unitarity that we
  discuss below.}. We therefore do not consider this operator further.

At dimension-4 in the SM factor there is the operator
\begin{align}
  i \left[ (D_\mu H)^\dagger D_\nu H - \text{h.c.} \right]
  \rightarrow m_Z(Z_\mu \partial_\nu h - Z_\nu \partial_\mu h),
  \label{eq:SMdim4}
\end{align}
concentrating on the part giving an $hZ$ interaction.
Including a DM factor, we consider
\begin{align}
  \frac{1}{\Lambda^3} \bar X \gamma^{\mu \nu} X \times i \left[ (D_\mu H)^\dagger D_\nu H - \text{h.c.} \right],
  \label{eq:dim7}
\end{align}
neglecting the similar possibility with $\bar X \gamma^5 \gamma^{\mu
  \nu} X$.

Finally there are dimension-5 SM operators \cite{Chen:2013gya}
\begin{align}
  ( & B_{\nu \mu} Y_H H^\dagger D^\nu H  + \text{h.c.} )  \label{eq:SMdim5a} \\
     &  \rightarrow  \frac{\langle v \rangle}{2} ( \cos \theta_w F_{\nu \mu} \partial^\nu h - \sin \theta_w Z_{\nu \mu} \partial^\nu h  ) \nonumber \\
  ( & W^a_{\nu \mu} H^\dagger t^a D^\nu H + \text{h.c.} )   \label{eq:SMdim5b} \\
     &  \rightarrow  - \frac{\langle v \rangle}{2} ( \sin \theta_w F_{\nu \mu} \partial^\nu h + \cos \theta_w Z_{\nu \mu} \partial^\nu h ) \nonumber
\end{align}
where $B_{\nu \mu}$, $W_{\nu \mu}^a$ are the field strengths for
$U(1)_Y$ and $SU(2)_L$, and $Z_{\nu \mu}, F_{\nu \mu}$ are the field
strengths for $Z$ and $\gamma$, respectively. Dimension-8 operators
are formed by including a DM factor of either $\bar{X}
\gamma^\mu X$ or $\bar{X} \gamma^{\mu 5} X$ for fermion DM, and either
$i(\phi^\dagger \partial^\mu \phi - h.c.)$ or $(\phi^\dagger
\partial^\mu \phi + h.c.)$ for scalar DM.  Combined with the
possibility of exchanging $B_{\nu \mu}, W^a_{\nu \mu}$ for $\tilde
B_{\nu \mu}, \tilde W^a_{\nu \mu}$, a large number of operators are
possible. 
We therefore restrict our attention to two representative examples
with scalar DM ($\phi$) or a Dirac fermion ($X$):
\begin{align}
&\frac{1}{\Lambda^4} \frac{1}{2} ( \phi^\dagger \partial^\mu \phi + \text{h.c.} ) \  ( B_{\nu \mu} H^\dagger D^\nu H + \text{h.c.} ) \label{eq:dim8scalar} \\
&\frac{1}{\Lambda^4} \bar{X} \gamma^\mu X \  ( W^a_{\nu \mu} H^\dagger t^a D^\nu H + \text{h.c.} ) \quad , \label{eq:dim8fermion}
\end{align}
Refs.~\cite{Chen:2013gya,Fedderke:2013pbc} discuss the complete list
of possible operators, as well as further details on the relic density
and gamma-ray signals of dark matter annihilation.

For the operators in Eqs.~(\ref{eq:dim7}), and
(\ref{eq:dim8scalar}-\ref{eq:dim8fermion}) we derive constraints on
$\Lambda$ as a function of DM mass from a mono-Higgs search. For LHC
Run 1 data, we consider the $b \bar b$ channel with the weakest cut on
the missing energy $\met > 120$ GeV. Higher $\met$ values will
necessarily require larger momentum transfer and thus lead to even
larger error in the validity of the EFT, as discussed further in the
following section. For 14 TeV, we obtain constraints using the
diphoton channel, where we find the best results.

The LHC Run 1 lower bounds on $\Lambda$ are comparable and on the order of
200 GeV for all three cases, increasing up to 300 GeV for 14 TeV
projections. The related operator $\frac{1}{\Lambda^4}\bar X
\gamma^\mu X (B_{\mu \nu}H^\dagger D^\nu H + \text{h.c.})$ was also
studied in Ref.~\cite{Carpenter:2013xra}, where they obtained very
similar bounds.

Even though one would expect the constraints on the dimension-7
operator to be stronger than for the dimension-8 ones, they are in
fact slightly weaker. This is because most of the mono-Higgs signal is
coming from the high momentum transfer ($\Qtr$) region, as can also be
seen in Fig.~\ref{fig:eft}, and the dimension-7 operator has a softer
$\Qtr$ dependence. This result is clearly related to the issue of
validity of the EFT, as we discuss further below.

\subsection{Unitarity}

A frequent concern in this EFT approach is that, taking LHC
constraints at face value, the values of $\Lambda$ that can be probed
correspond to energy scales accessible at the LHC. This implies a
violation of perturbative unitarity at high momentum transfer, or
equivalently that the EFT is no longer a valid description for LHC
processes.

Fig.~\ref{fig:eft} shows the distribution for the momentum transfer
$\Qtr$ for the operator of Eq.~(\ref{eq:dim8fermion}). Compared to the
naive constraint of $\Lambda \gtrsim 225\ \GeV$ derived for the
operator, it is clear that the EFT description is on shaky footing.
For an $s$-channel mediator, the condition $\Qtr \lesssim 4 \pi
\Lambda$ is required for an expansion in the mediator mass for a
perturbative theory \cite{Busoni:2013lha} or $\Qtr \lesssim 2.5
\Lambda $ for unitarity of the S-matrix \cite{Shoemaker:2011vi}.  In
general the specific regime of $\Qtr$ where the theory breaks down
depends on the form of the operator (as well as its UV
completion). Since it is not straightforward to derive UV completions
for the operators here, we consider $\Qtr = 4 \Lambda$ and $\Qtr =
4\pi \Lambda $ as representative of where the EFT assumption begins to
suffer from large errors.

We implement three different criteria: $Q_{\text{tr}} < \infty$
(corresponding to the na\"{i}ve limit), $Q_{\text{tr}} < 4\pi \times
\Lambda$, and $Q_{\text{tr}} < 4 \Lambda$. More specifically, for a
given $\Lambda$, we discard any events in violation and thus rescale
the calculated cross section by the fraction of events satisfying this
criterion at parton-level. The conditions above on the generated
events should not be taken literally; they are only to indicate the
size of the error in assuming a single effective operator can describe
the relevant physics. This procedure gives conservative constraints,
in the sense that any new physics giving rise to the operator is
expected to be relevant at these scales. In general, this could lead
to even stronger constraints on the model, for example from an
enhanced signal in the original channel or from other new signal
channels \cite{Busoni:2013lha,Shoemaker:2011vi,Buchmueller:2013dya}.

Our results for the operators are shown in Fig.~\ref{fig:eftbounds},
where the solid lines give the lower limit on $\Lambda$ without any
condition on the momentum transfer. When a condition on $\Qtr$ is
imposed, this weakens and shifts the bound on $\Lambda$; in
addition, low values of $\Lambda$ are no longer excluded, which we
interpret as the breakdown of the EFT. This is also illustrated by
Fig.~\ref{fig:eft}, where we show the mono-Higgs cross section when
each one of the unitarity conditions above is imposed. For very small
$\Lambda$, no events satisfy the condition on $\Qtr$. As $\Lambda$ is
increased, more events meet the criterion until the suppression of the
cross section with large $\Lambda$ takes over. The excluded region is
the range of $\Lambda$ where the cross section is above that
observable at the LHC (indicated by the dashed line).

For the weakest condition $Q_{\text{tr}} < 4\pi \Lambda$ a constraint
is possible for all operators below DM masses around a few hundred
GeV. In the most restrictive case $Q_{\text{tr}} < 4 \Lambda$, we find
that no bound is possible for the operators in
Eqs.~(\ref{eq:dim7},\ref{eq:dim8scalar}). For the fermion DM operator
in Eq.~(\ref{eq:dim8fermion}), a limit for a narrow range in $\Lambda$
is still possible with the strongest $\Qtr$ condition and 8 TeV data,
but again no bound is expected at larger masses or with a 14 TeV
run. Compared to the results for the 7/8 TeV runs of the LHC, the 14
TeV run does not necessarily promise a significant improvement with
respect to the issue of unitarity due to the need for a stronger
$\met$ cut to suppress backgrounds.

\section{Dark matter via a $Z'$ and heavy Higgs}
\label{sec:ZPrime}

Motivated by the mono-Higgs processes discussed in the previous
section, we construct a simple model with renormalizable interactions
where the relevant states may be produced on-shell. The high-dimension
operators considered previously are challenging to UV-complete;
however, it is more straightforward to generalize the mono-Higgs
process, as shown in Fig.~\ref{fig:diagrams}. If the intermediate $Z$
is instead a new $Z'$ gauge boson, resonant production is possible;
the $Z'$ then decays to a Higgs plus an intermediate state which
decays to a DM pair. Since a SM state decaying to DM is highly constrained,
 we consider a two-Higgs doublet extension to the standard
model with $Z' \to h A^0$, where $A^0$ is a heavy pseudoscalar with
a large branching ratio to dark matter. Below we discuss in more
detail the $Z'$ coupled to a two-Higgs doublet model (2HDM), which is
sufficient to determine the mono-Higgs signal. More model-dependent
details of the DM coupling to the pseudoscalar are discussed in
Sec.~\ref{sec:DMcoupling}.

The gauge symmetry of the SM is extended by a $U(1)_{Z'}$, with a new
massive $Z'$ gauge boson (see, for example,
\cite{Langacker:2008yv,Carena:2004xs}). We assume that this sector
also contains a SM singlet scalar $\phi$ that leads to spontaneous
breaking of the symmetry and a $Z'$ mass at a scale above electroweak
symmetry-breaking. There are many choices for how the SM fermions are
charged under the $U(1)_{Z'}$; for simplicity, we assume
generation-independent charges for the fermions and that only the
right-handed quarks $u_R$ are charged\footnote{ Anomaly cancellation
  can be achieved with a pair of colored triplet fields which are
  singlets with respect to $SU(2)_L$: $\psi_L (Q_z=0,Y=-2/3)$ and
  $\psi_R (Q_z=-z_u, Y=-2/3)$ where $z_u$ is the $Z'$ charge of $u_R$.
  }. 
This allows LHC production of the $Z'$, but since the leptons are
neutral, avoids potentially stringent constraints from searches for
dilepton resonances.

For the Higgs sector we assume a Type 2 two-Higgs-doublet model, where
$\Phi_u$ couples to up-type quarks and $\Phi_d$ couples to down-type
quarks and leptons:
\begin{equation}
  -{\cal L} \supset  y_u Q \tilde \Phi_u \bar u + y_d Q \Phi_d \bar d + y_e L \Phi_d \bar e  + {\rm h.c.}
\end{equation}
with hypercharge $Y=1/2$ Higgs doublets $\Phi_u, \Phi_d$ that could
have $Z'$ charges $z_u, z_d$. In the case we consider, only $u_R$ and
$\Phi_u$ are charged under $U(1)_{Z'}$. Our convention for the charges
are shown in Table~\ref{tab:charges}.

After electroweak symmetry breaking, the Higgs doublets attain vevs
$v_u$ and $v_d$, and in unitary gauge the doublets are parametrized as
\begin{align}
\Phi_d &= \frac{1}{\sqrt{2}}
\begin{pmatrix}
-\sin{\beta} \ H^+ \\ v_d - \sin{\alpha} \ h + \cos{\alpha} \ H - i \sin{\beta} \ A^0
\end{pmatrix} 
\quad , \nonumber \\
\Phi_u &= \frac{1}{\sqrt{2}}
\begin{pmatrix}
\cos{\beta} \ H^+ \\ v_u + \cos{\alpha} \ h + \sin{\alpha} \ H + i \cos{\beta} \ A^0
\end{pmatrix}
\end{align}
where $h,H$ are neutral CP-even scalars and $A^0$ is a neutral CP-odd
scalar. Furthermore, $\tan{\beta} \equiv v_u/v_d$, and $\alpha$ is the
mixing angle that diagonalizes the $h - H$ mass squared matrix.

We make some simplifying assumptions for the Higgs sector, taking $h$
as the scalar corresponding to the observed Higgs boson with $m_h \sim
125$ GeV. The remaining scalars $H, A^0, H^\pm$ are assumed to have
masses around or above $300$ GeV, in accordance with $b \to s \gamma$
constraints \cite{Branco:2011iw}.  Fits to the observed Higgs
couplings from the LHC \cite{Craig:2013hca} indicate that a Type 2
2HDM is tightly constrained around the alignment limit where
$\sin{(\beta-\alpha)} \rightarrow 1$ (specifically $\beta \rightarrow
\alpha + \pi/2$, $\alpha \in ( -\pi/2, 0)$). In this limit, $h$ has
SM-like couplings to fermions and gauge bosons. In addition,
perturbativity of the top yukawa coupling implies $\tan{\beta} \gtrsim
0.3$. Hence, we choose to work in the $\alpha - \beta$ parameter space
where $\tan{\beta} \ge 0.3$ and $\alpha = \beta - \pi/2$.

\renewcommand\arraystretch{1.4}
\begin{table}[t]
\begin{tabular}{| >{\centering\arraybackslash}m{.5in} | >{\centering\arraybackslash}m{.5in} | >{\centering\arraybackslash}m{.5in} |
 >{\centering\arraybackslash}m{.5in} | >{\centering\arraybackslash}m{.5in} |
 >{\centering\arraybackslash}m{.5in} |}
    \hline
\centering
     & $\Phi_d$ & $\Phi_u$ & $Q_L$ & $d_R$ &  $u_R$  \\ \hline \hline
    $U(1)_{Z'}$  & 0  & 1/2 & 0 & 0 & 1/2 \\ \hline
     \end{tabular}
\caption{SM fermion and scalar $U(1)_{Z'}$ gauge charges. All other
  SM particles are neutral.
}
\label{tab:charges}
\end{table}
\renewcommand\arraystretch{1.0}

The Higgs vevs lead to $Z-Z'$ mass mixing. Diagonalizing the gauge
boson mass matrix, the tree-level masses of the $Z$ and $Z'$ bosons are given by
\begin{align}
  M_Z^2 &\approx ( M_{Z}^0)^2  - \epsilon^2 \left[( M_{Z'}^0)^2 - ( M_{Z}^0)^2 \right] \nonumber \\
  M_{Z'}^2 &\approx ( M_{Z'}^0)^2 + \epsilon^2 \left[( M_{Z'}^0)^2 - ( M_{Z}^0)^2 \right]
\quad ,
\label{eq:Zpmasses}
\end{align}
where $(M_Z^0)^2 = g^2(v_d^2+ v_u^2)/(4\cos^2{\theta_w}) $ and
$(M_{Z'}^0)^2 = g_z^2 ( z_d^2 v_d^2 + z_u^2 v_u^2 + z_\phi^2
v_\phi^2)$ are the mass-squared values in the absence of mixing. The
result above is accurate to order $\epsilon^2$, where $\epsilon$ is
a small mixing parameter given by
\begin{align}
	\epsilon & \equiv \frac{1}{M_{Z'}^2 - M_Z^2} \frac{g g_z}{2 \cos{\theta_w}} ( z_d v_d^2 + z_u v_u^2) \nonumber \\
	& =  \frac{(M_Z^0)^2}{M_{Z'}^2 - M_Z^2} \frac{2 g_z \cos \theta_w}{g}  z_u \sin^2 \beta.
	\label{eq:epsilon}
\quad
\end{align}
Finally, the mass eigenstates corresponding to the observed $Z$ boson
and the hypothetical $Z'$ boson are
\begin{align}
  Z^\mu &\approx W^{3 \mu} \cos{\theta_w} - B_Y^\mu \sin{\theta_w} + \epsilon B_Z^\mu \quad , \nonumber \\
  Z'^\mu &\approx B_Z^\mu - \epsilon \left( W^{3\mu} \cos{\theta_w} - B_Y^\mu \sin{\theta_w}  \right).
  \quad 
\end{align}

\subsection{$Z'$ constraints}

\begin{figure}[t]
\mbox{\includegraphics[width=0.49\textwidth,clip]{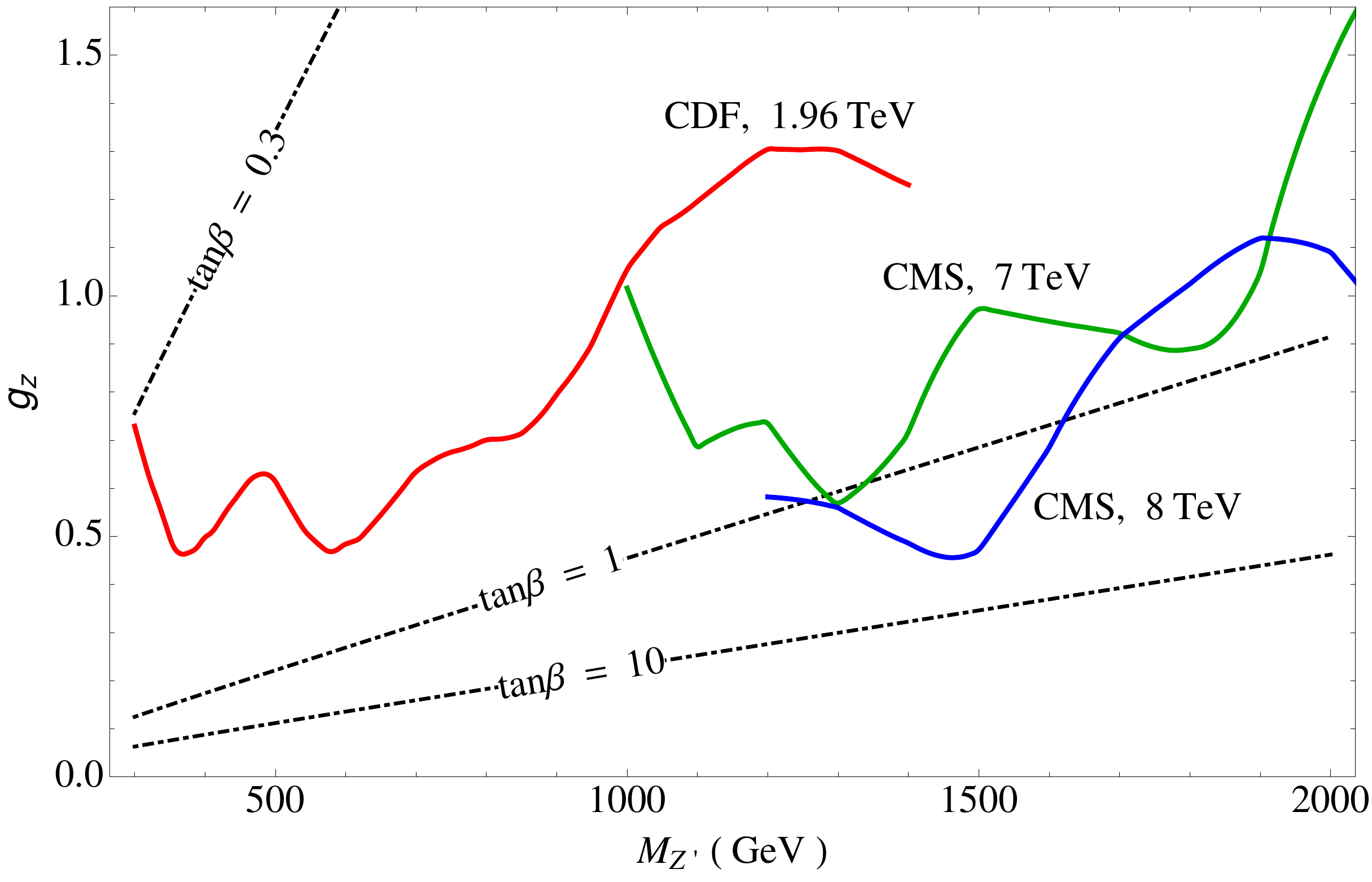}} 
\caption{95$\%$ CL bounds on the $Z'$ coupling $g_z$ as a function of
  $M_{Z'}$. The dashed lines are upper bounds from $\rho_0$ parameter
  constraints on $Z-Z'$ mixing, given in Eq.~(\ref{eq:rhoconstraint}),
  for three values of $\tan \beta = 0.3, 1, 10$. We also show upper
  limits from dijet resonance searches at the Tevatron and at the LHC;
  see text for further details.
\label{fig:gzlimits}}
\end{figure}

The $Z-Z'$ mixing leads to a modification to the $Z$ mass, as shown in
Eq.~(\ref{eq:Zpmasses}). This in turn affects the relation between the
$W$ and $Z$ masses, which is expressed as a deviation of the $\rho_0$
parameter away from unity:
 \begin{align}
   \rho_0 = 1 + \epsilon^2 \left(  \frac{M_{Z'}^2 - M_Z^2}{ M_Z^2} \right)
   \quad ,
\label{eq:rho}
\end{align}
Current precision electroweak global fits constrain $\rho_0 =
1.0004^{+0.0003}_{-0.0004}$ \cite{Beringer:1900zz}.  Taking this
result at face value, the approximate 95$\%$ upper limit
\begin{equation}
	\rho_0  \leq 1.0009
	\label{eq:rhoconstraint}
\end{equation} 
implies an upper limit on $g_z$ (at fixed $\tan \beta$ and $M_{Z'}$),
shown in Fig.~\ref{fig:gzlimits}.  Although there are also strong LEP
constraints on $Z-Z'$ mixing from, {\it e.g.}, precision measurements
of dijets/dileptons through a $Z$ resonance, in our case the limits
are weak since the coupling to leptons is $\epsilon$-suppressed.

There are additional $g_z$ constraints from searches for dijet
resonances from $Z'$ decay to $q \bar q$, also shown in
Fig.~\ref{fig:gzlimits}. We apply results from Tevatron and LHC
studies, with Tevatron results \cite{Aaltonen:2008dn} providing
coverage for $300\ \GeV \leq M_{Z'} \leq 1.4$ TeV.  We also apply
95$\%$ CL upper limits from CMS using 7 TeV \cite{CMS:2012yf} and 8
TeV \cite{CMS-PAS-EXO-12-059} data\footnote{The ATLAS collaboration
  has also presented 95$\%$ CL upper limits
  \cite{Aad:2011fq,ATLAS-CONF-2012-148}, but for a narrow Gaussian in
  dijet mass distribution, which is not applicable to this case since
  there is a significant tail to the distribution at lower dijet
  masses.}, given in a model-independent form in terms of a cross
section times acceptance for a narrow resonance decaying to $q \bar
q$. An upper bound on $g_z$ is derived by comparing our detector-level
simulation to the published upper limits, assuming that the $Z'$ width
is fixed for the most part by its decay to quarks:
\begin{equation}
	\Gamma_{Z' \to q \bar q} \approx \frac{g_z^2}{24\pi} z_u^2 N_c M_{Z'}
\end{equation}
for each light-quark flavor.  This is a valid approximation for the
model here, assuming that there isn't a significant width for $Z'$ decay to
other new fermionic modes.

For masses below $\sim 1.3$ TeV (exactly the regime that we find the
strongest potential mono-Higgs signal) and in particular for large
$\tan \beta$, we find that the $\rho_0$ constraint on $g_z$ is
stronger than dijet limits. However, for $\tan \beta \lesssim 0.6$,
the dijet constraints dominate even at low masses. For the remainder
of the paper, for any given $M_{Z'}$ and $\tan \beta$, we will simply
assume the coupling $g_z$ is the maximum allowed by $\rho_0$ and dijet
constraints, as given in Fig.~\ref{fig:gzlimits}.

\subsection{Mono-Higgs signal}


\begin{figure*}[t]
\mbox{\includegraphics[width=0.49\textwidth,clip]{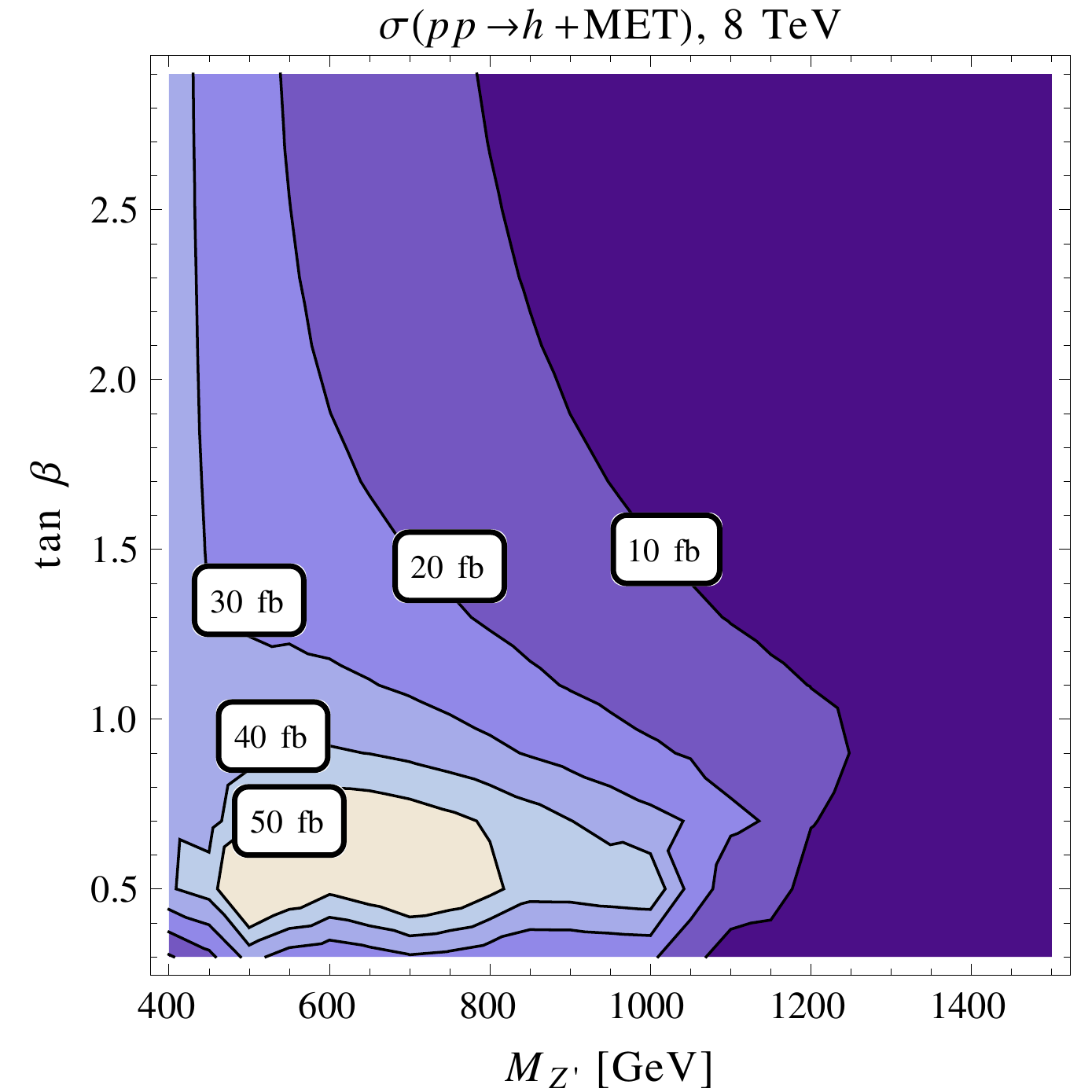}} 
\mbox{\includegraphics[width=0.49\textwidth,clip]{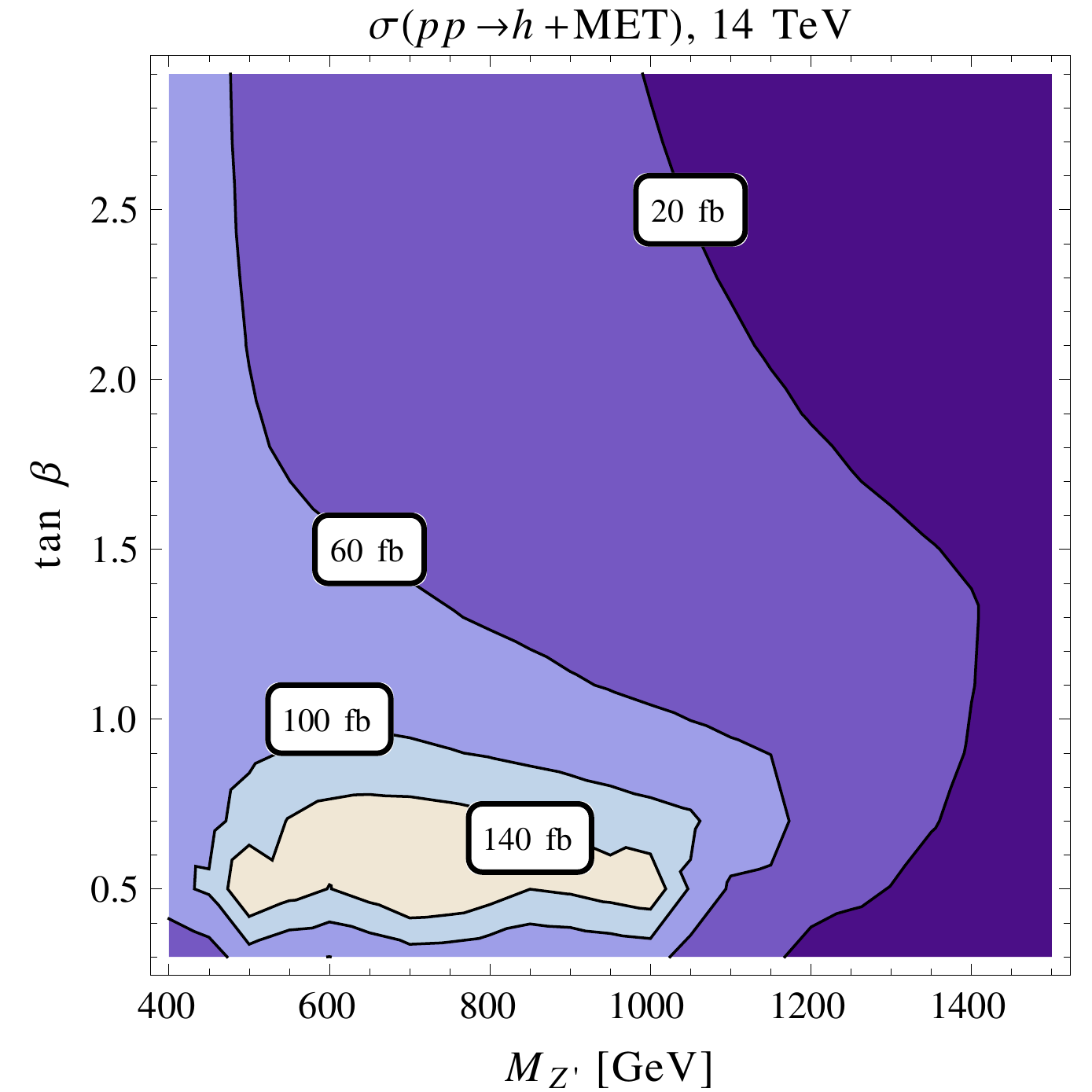}} 
\caption{Total cross sections for Higgs+MET, via a new $Z'$ gauge
  boson coupled to a 2HDM, for the LHC at 8 TeV and 14 TeV. Both Higgs
  plus DM production from $Z' \to h A^0$ and Higgs plus MET from $Z'
  \to h Z, Z \to \nu \bar \nu$ are included. The $Z'$ gauge coupling
  is fixed to be its 95$\%$ CL upper limit, as shown in
  Fig.~\ref{fig:gzlimits}.}
\label{fig:xsec}
\end{figure*}


The mono-Higgs signal associated with DM plus Higgs production
proceeds through $Z' \to hA^0$; the decay width for this
 to leading order in $\epsilon$ is
\begin{equation}
   \Gamma_{Z' \to hA^0} =  (g_z \cos \alpha \cos \beta)^2 \frac{|p|}{24 \pi} \frac{|p|^2}{M_{Z'}^2}.
\end{equation}
The center of mass momentum for the decay products is $|p| =
\frac{1}{2 M_{Z'}} \lambda^{1/2}(M_{Z'}^2,m_h^2, m_{A^0}^2)$, where
$\lambda$ is the K\"{a}llen triangle function.  Since only the
$\Phi_u$ doublet couples directly to the $Z'$, and since the
pseudoscalar component of the $\Phi_u$ scales with $\cos \beta$, this
decay width is suppressed by $1/\tan^2 \beta$ in the limit of large
$\tan \beta$. For $\tan \beta < 1$, the rate actually increases
because the allowed $g_z$ from the precision electroweak constraint
increases, at least until $\tan \beta \approx 0.6$ when dijet limits
take over.

The $Z'$ model enjoys an additional source of Higgs plus MET from the
decay of $Z' \to h Z$, where the $Z$ decays invisibly. The decay width
is
\begin{equation}
   \Gamma_{Z' \to hZ}  = (g_z \cos \alpha \sin \beta)^2 \frac{|p|}{24 \pi} \left( \frac{ |p|^2 }{M_{Z'}^2} + 3 \frac{M_Z^2}{M_{Z'}^2} \right),
\end{equation}
which grows with smaller $M_{Z'}$ due to the $M_Z^2/M_Z'^2$ term.
At fixed $M_{Z'}$, the mono-Higgs rate for this process is almost
independent of $\tan \beta$ for $\tan \beta \gtrsim 0.6$.  Although
the rate na\"{\i}vely scales as $\sin^4 \beta$, this dependence is
almost exactly cancelled when we apply the upper limit on $g_z$ from
$\rho_0$, which leads to an upper limit on $g_z \propto 1/(\sin^2
\beta)$. This can also be seen from
Eqs.~(\ref{eq:epsilon},\ref{eq:rho}). When $\tan \beta \lesssim 0.6$,
the constraint on $g_z$ is independent of $\tan \beta$ and the width
is therefore suppressed by $\sin^4 \beta$.

\begin{figure}[tbh]
\mbox{\includegraphics[width=0.49\textwidth,clip]{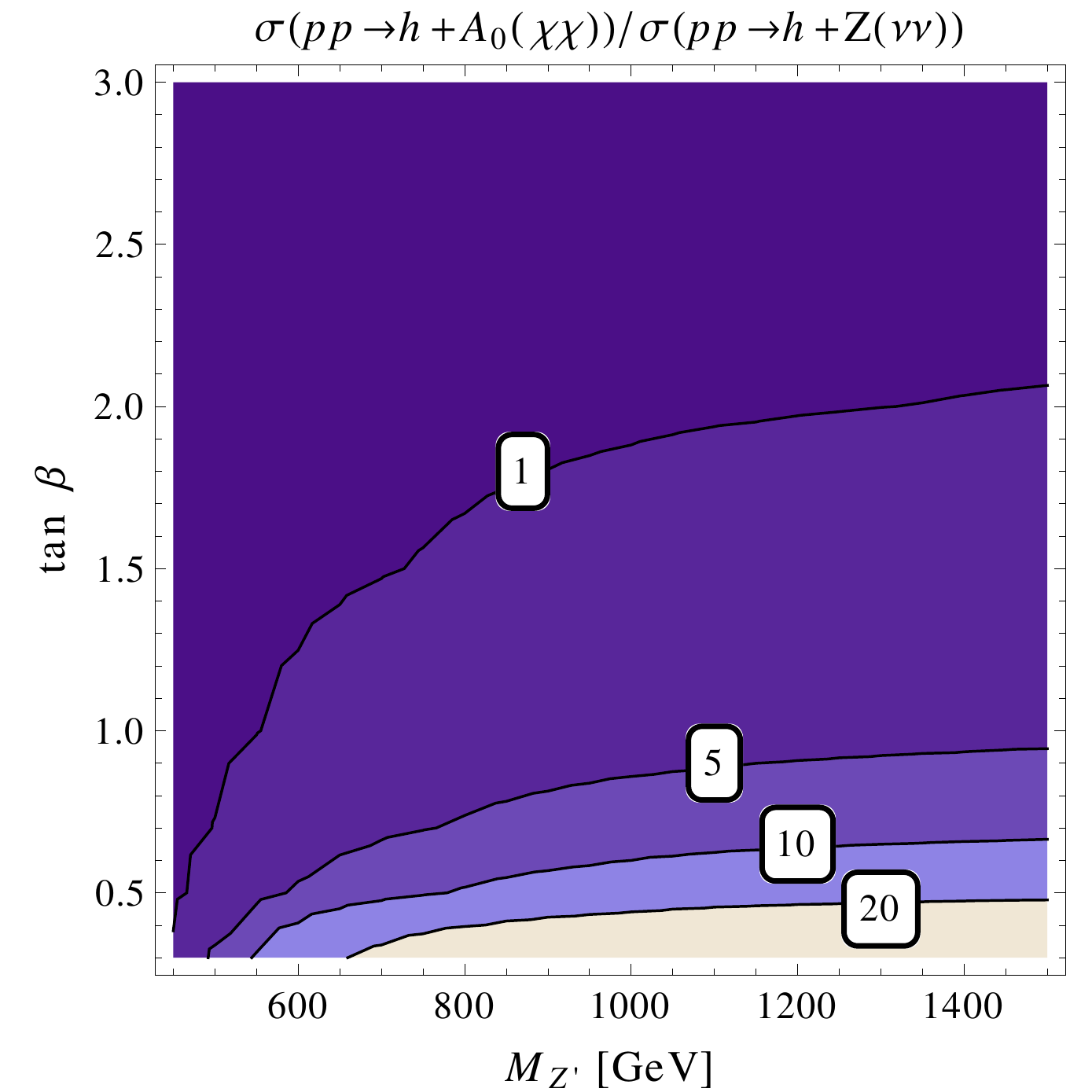}} 
\caption{Ratio of the cross sections (or, ratio of the
  branching ratios) for mono-Higgs from $Z' \to hA^0, A^0 \to \bar X
  X$ to mono-Higgs from $Z' \to hZ, Z \to \bar \nu \nu$.}
\label{fig:xsecratio}
\end{figure}

Fig.~\ref{fig:xsec} shows the total mono-Higgs cross section at 8 TeV
and at 14 TeV, as a function of $M_{Z'}$ and $\tan \beta$.  We have
fixed the coupling $g_z$ according to its 95$\%$ CL upper bound, as
discussed in the previous section. 
The heavy scalar masses are assumed to be 300 GeV and we take the
alignment limit, $\sin(\beta - \alpha) = 1$. The branching ratio of
$A^0$ to dark matter is taken to be 100$\%$.  Despite the larger
coupling allowed at larger $M_{Z'}$, the total cross section
eventually falls with $M_{Z'}$ due to pdf suppression. For large or
small $\tan \beta$, the cross section also falls due to the $(\sin
\beta \cos \beta)^2$ dependence in the $hA^0$ channel. The ratio of
the two mono-Higgs rates is shown in Fig.~\ref{fig:xsecratio}. Over
much of the parameter space we consider, the mono-Higgs from $Z' \to
hA^0$ dominates, however $Z' \to hZ$ is a non-negligible fraction of
the total signal and becomes important at low $M_{Z'}$ and also at
large $\tan \beta$.

\begin{figure}[t]
\mbox{\includegraphics[width=0.49\textwidth,clip]{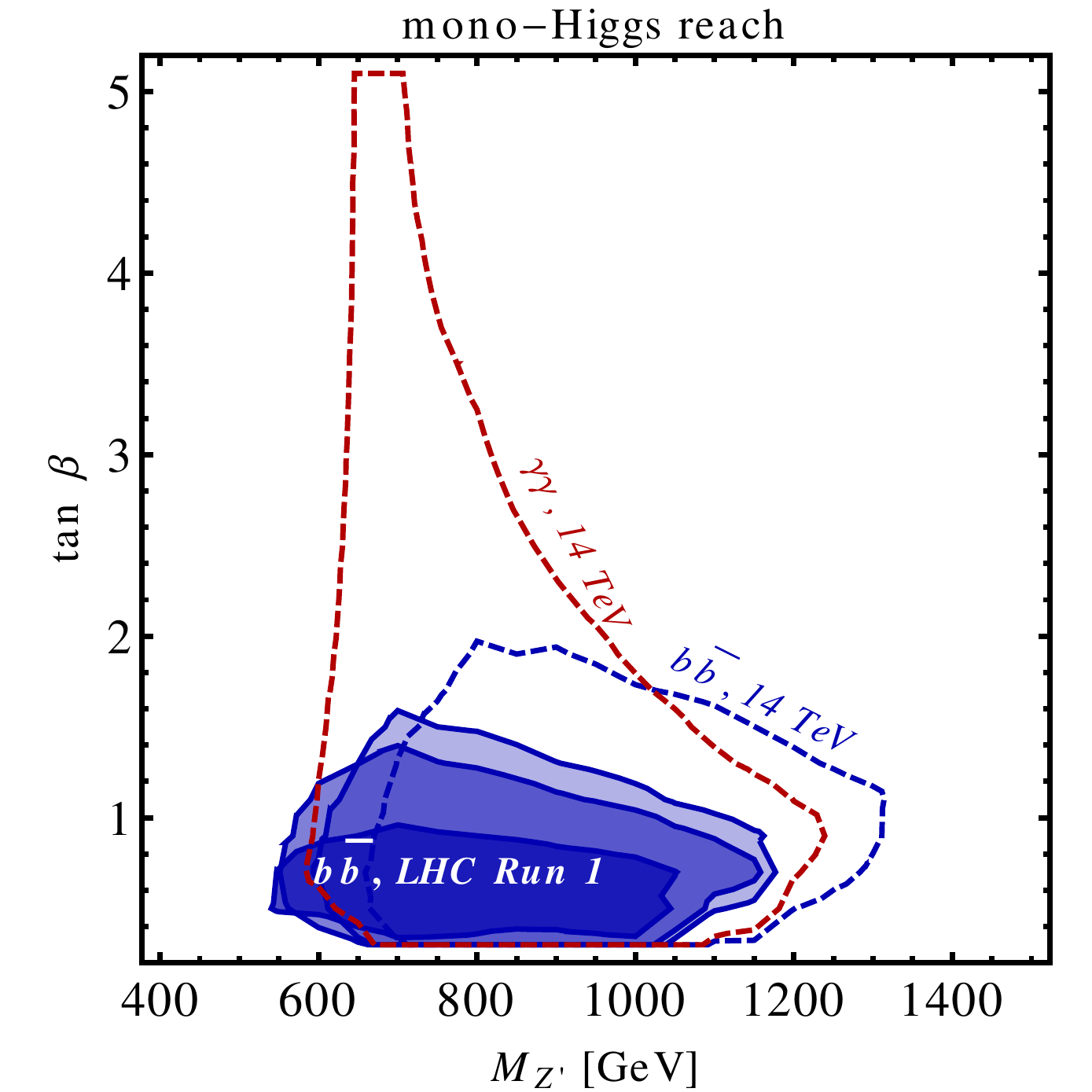}} 
\caption{95$\%$ CL exclusion regions for the parameter space with data
  from Run 1 of the LHC (7 and 8 TeV, total 25/fb) for the $b \bar b$
  channel with MET cuts of 120, 160 and 200 GeV. Dashed lines give
  projections for a 14 TeV LHC with 300/fb integrated luminosity for
  $b \bar b$ and diphoton channels. We only show the parameter space
  up to $\tan \beta = 5$ but the reach for the diphoton channel could
  extend to somewhat larger $\tan \beta$, since the cross section is
  approximately constant with $\tan \beta$.}
\label{fig:LHCreach}
\end{figure}

We present results for the mono-Higgs reach at the LHC in
Fig.~\ref{fig:LHCreach}. For Run 1 of the LHC (combined 7 TeV and 8
TeV), we show the three 95$\%$ CL exclusion regions for the $b \bar b$
channel with $\met > 120,160,$ and 200 GeV, where the constrained
region increases with MET cut\footnote{ If we were to use the results
  of Ref.~\cite{Carpenter:2013xra}, the 8 TeV data would be
  unconstraining at 95$\%$ CL for almost the entire parameter
  space. This is partly due to the rather conservatives estimates and
  also because the cuts are not optimal for our model.}. For 14 TeV
projections, we again find better overall sensitivity with a harder
MET cut (taken here to be $\met > 250$ GeV) to reduce SM backgrounds.

The diphoton channel is sensitive to lower cross sections compared to
$b\bar b$ for a 14 TeV LHC, as evidenced by the reach of this channel
for large values of $\tan \beta$. Although our plot cuts off at $\tan
\beta =5$, the mono-Higgs cross section is approximately constant for
large $\tan \beta$ and the sensitivity can extend to much higher $\tan
\beta$. However for much larger $\tan \beta$, direct searches for $H,
A^0$ would start to be constraining \cite{CMS-PAS-HIG-13-021},
depending on the scalar masses. The diphoton channel also performs
worse than expected at large $M_{Z'}$. This is because in our detector
simulation, the energy resolution for photons deteriorates at higher
energies such that the $m_{\gamma\gamma}$ peak is much broader, which
limits the signal efficiency. This effect could be reduced by
loosening the cut on $m_{\gamma\gamma}$, however the extent to which
this would be helpful depends on the actual energy resolution in the
experiment.

An appropriate question is whether other 14 TeV searches will
potentially also have sensitivity for this model. For example,
although data from the next LHC run will improve dijet resonance
constraints, this will be mainly at large $M_{Z'}$; below 1.5 TeV it
will be even more difficult to probe due to the large QCD backgrounds.
Here the strongest constraint for our model was the precision
electroweak fit for $\rho_0$. A somewhat indirect but possibly
important channel is a direct search for $H,A^0$ decay; for example,
for $H$ decay to SM fermions, the 14 TeV data could improve the upper
limits on $\tan \beta$ significantly for the range of masses relevant
here \cite{Arbey:2013jla}.

\subsection{Dark Matter Coupling to Higgs Sector \label{sec:DMcoupling}}

To incorporate DM interactions, we have assumed that the CP-odd
pseudoscalar $A^0$ of the theory possesses a large coupling to DM
particles, such that the branching ratio is order one. Here we sketch
out some simple models that could give rise to this kind of coupling,
reserving more detailed studies for future work.

One possibility is fermion DM; for example, a pseudoscalar interaction
can arise in singlet-doublet DM from a coupling to the down-type
Higgs.  In this model, a singlet $S$ and electroweak doublets
$D_{1,2}$ (all singlets under $U(1)_{Z'}$) are introduced, with a
Lagrangian
\begin{align}
	-{\cal L} \supset \frac{1}{2} M_S^2 S^2 + M_D D_1 D_2 +
	    y_1 S D_1 \Phi_d  + y_2 S \Phi_d^\dagger D_2 + \text{h.c.} \nonumber
\end{align}
The DM is the Majorana fermion that is the lightest mass eigenstate,
and we require that it has a mass of at least $m_h/2$ in order to
avoid bounds on the invisible width of the Higgs. In general, this
state is a mixture of the singlet and the neutral components of the
doublets.  For more details, see for example
Refs.~\cite{Cohen:2011ec,Cheung:2013dua}.

Elastic scattering off quarks can proceed via the exchange of $h$ or
$H$, and direct detection constraints severely restrict the parameter
space for this model. However, in parts of the parameter space near
the ``blind spot'' where the coupling through the Higgs is suppressed,
the direct detection cross sections are small. This cancellation
requires $\tan \theta < 0$, where $y_1 = y \cos \theta, y_2 = y \sin
\theta$.  We find it is possible to obtain large branching ratios of
$A^0$ to DM while satisfying LUX constraints \cite{Akerib:2013tjd} for
parameter values of $y = 1.5$, $\tan \theta = -2$ and masses of $M_S
\approx 100-200$ GeV and $M_D \approx 120-180$ GeV.

For scalar DM, we consider a complex scalar field $X$, written as $X =
\frac{1}{\sqrt{2}} \left( X_1 + i X_2 \right)$, which is a SM singlet
and has $U(1)_{Z'}$ charge $-1/4$. Then the renormalizable
interactions of the DM with the Higgs sector are
\begin{align}
\mathcal{L} \supset & \ \left( \lambda_{\text{dd}} |\Phi_d|^2 + \lambda_{\text{uu}} |\Phi_u|^2 \right) | X |^2 \nonumber \\
   & \ \ \ \ +  \left( \lambda_{\text{du}} \Phi_d^\dagger \Phi_u X^2 + \text{h.c.} \right),
\end{align}
with all couplings taken to be real.
The mass eigenstates are the real fields fields $X_{1,2}$ with masses
$ m_{1,2}^2 = m_X^2 \mp 2 \lambda_{\text{du}} \sin(2\beta) m_W^2/g^2$,
where the overall mass scale $m_X^2$ is a free parameter.  Again, the
lightest component is a DM candidate.

The $A^0$ can decay through the term $\lambda_{\text{du}} v A^0 X_1
X_2$. However, this decay is not truly invisible, since the $X_2$ can
decay to $X_1 q \bar q$ through an off-shell $A$ or $Z'$, as well as
to $X_1 \ell^+ \ell^-$ with a somewhat smaller rate. This $X_2$ decay
will wash out some of the missing energy; however, if the splitting
between $X_2$ and $X_1$ is not too large, these additional jets or
leptons are relatively soft. There is some tension for this parameter
space, since larger $\lambda_{\text{du}}$ is needed for an ${\cal
  O}(1)$ branching fraction, but at the same time this leads to a
larger mass splitting.

Finally, DM scattering off of quarks is through $h$ or $H$ exchange,
since $Z'$ interactions are inelastic with a large mass splitting.
It is possible to satisfy the direct detection limits from LUX if
there are cancellations among the couplings $\lambda_{\text{dd}}$,
$\lambda_{\text{uu}}$, and $\lambda_{\text{du}}$ at the 10$\%$ level
\cite{He:2013suk}. We find that couplings of order $|\lambda| \sim
0.1$ and a mass scale of $m_X \sim 100$ GeV can give rise to the
desired features of the model.

\section{Summary and Conclusions}
\label{conc}

The discovery of a new particle brings with it the prospect of a new
signal channel for probing dark matter particle physics. In the search
for dark matter, there are already many different potential avenues to
its discovery, though so far without conclusive results.  The simple
question motivating this work is to search for possible models where
dark matter production with a Higgs is the dominant discovery mode in
the current generation of hadron colliders.  For these models we
adopted ATLAS results from the combined 7 and 8 TeV (25/fb) analysis
in the $h \to \bar b b$ channel in order to derive constraints, and
studied the sensitivity of a 14 TeV LHC in the $\bar b b$ as well as
diphoton channels.

One way for mono-Higgs to occur is through higher dimension operators
coupling dark matter to Higgs doublets and electroweak gauge
bosons. LHC constraints applied to the dimension-7 or -8 operators
studied here lead to the na\"{\i}ve conclusion that the cutoff scale
$\Lambda$ must be greater than 100-200 GeV. However, this is
problematic from an effective field theory point of view, since such
scales are low compared to the typical momentum transfer in the
collider process. We have attempted to quantify the extent to which
imposing a unitary constraint gives rise to a reliable (although
conservative) bound on the operator. This is possible only
for low dark matter masses.

We also presented a viable simplified model, where the resonant
production of a $Z'$ decaying to $h A^0$ and $hZ$ allows for a
potentially observable rate of mono-Higgs. This is primarily possible
at low $\tan \beta$, in part because the $Z' \to h A^0$ branching
fraction is 1/$\tan \beta^2$ suppressed. In addition, we require the
$Z'$ gauge coupling to be near the maximum allowed from precision
electroweak fits. Nevertheless, we show there is an interesting part
of parameter space for low $\tan \beta$ and $M_{Z'}$ around 1 TeV,
assuming the pseudoscalar $A^0$ decays to dark matter 100$\%$ of the
time.  We briefly discussed possible models that could give rise to
this large pseudoscalar to invisible branching ratio.  It would be
interesting to pursue more detailed model-building work in this
direction, taking into account direct detection or relic density
considerations.

\emph{Acknowledgments}: We would like to thank Dan Hooper, Austin
Joyce, Bj\"{o}rn Penning, Sean Tulin, and Daniel Whiteson for valuable
discussion.  L.T.W. is supported by a DOE Early Career Award under
grant DE-SC0003930.  This work was supported in part by the Kavli
Institute for Cosmological Physics at the University of Chicago
through grant NSF PHY-1125897 and an endowment from the Kavli
Foundation and its founder Fred Kavli.

\bibliography{monohiggs}

\end{document}